\documentclass[11pt]{article}
\usepackage{graphicx}
\usepackage{authblk}
\usepackage[textwidth=4.5cm]{geometry}
\usepackage{blindtext}

\usepackage{indentfirst,csquotes}

\usepackage{amssymb,amsthm,amsmath}
\usepackage{xcolor,paralist,hyperref,titlesec,fancyhdr,etoolbox}
\usepackage{graphicx}
\usepackage{tabularx}
\usepackage{longtable}

\newcolumntype{C}{>{\centering\arraybackslash}X}
\usepackage{caption}
\usepackage{subcaption}
\usepackage{multirow}
\usepackage{colortbl} 
\usepackage{makecell} 

\usepackage{lipsum}

\begin{document}
\title{
Inferring diagnostic and prognostic gene expression signatures across WHO glioma classifications: A network-based approach}

\author[1,2,*]{Roberta Coletti}
\author[3]{ M\'{o}nica L. Mendon\c{c}a}
\author[3,4]{Susana Vinga}
\author[1,2,5,*]{ Marta B. Lopes}
\affil[1]{ NOVA School of Science and Technology (FCT NOVA), University of Lisbon, Caparica, Portugal}
\affil[2]{  Center for Mathematics and Applications (NOVA MATH), FCT NOVA, University of Lisbon, Caparica, Portugal } 
\affil[3]{ INESC-ID, Instituto Superior Técnico, University of Lisbon, Lisbon, Portugal}
\affil[4]{IDMEC, Instituto Superior Técnico, University of Lisbon, Lisbon, Portugal}
\affil[5]{ UNIDEMI, Department of Mechanical and Industrial Engineering, NOVA SST, NOVA University of Lisbon, Lisbon, Portugal}
\affil[*]{{ Corresponding author. Marta B. Lopes: marta.lopes@fct.unl.pt ; Roberta Coletti: roberta.coletti@fct.unl.pt}}
\maketitle


\begin{abstract} 
Tumor heterogeneity is a challenge to designing effective and targeted therapies. Glioma-type identification depends on specific molecular and histological features, which are defined by the official WHO classification CNS. These guidelines are constantly updated to support the diagnosis process, which affects all the successive clinical decisions. In this context, the search for new potential diagnostic and prognostic targets, characteristic of each glioma type, is crucial to support the development of novel therapies. Based on a TCGA glioma RNA-sequencing dataset updated according to the 2016 and 2021 WHO guidelines, we proposed a two-step variable selection approach for biomarker discovery. Our framework encompasses the graphical lasso algorithm to estimate sparse networks of genes carrying diagnostic information. These networks are then used as input for regularised Cox survival regression model, allowing the identification of a smaller subset of genes with prognostic value. In each step, the results derived from the 2016 and 2021 classes were discussed and compared. For both WHO glioma classifications, our analysis identifies potential biomarkers, characteristic of each glioma type. Yet, better results were obtained for the WHO CNS classification in 2021, thereby supporting recent efforts to include molecular data on glioma classification.
\end{abstract}
\bigskip
{\textbf{Keywords:}
Graphical Lasso, Regularised Cox regression, Survival analysis, RNASeq, Glioma, TCGA, WHO CNS,  Networks, Variable selection , Biomarkers}
\bigskip

\section{Introduction}

Glioma is one of the most common brain tumors, which in adults represents more than $80\%$ of malignant cancers in this organ \cite{Ostrom2021}. 
One of the reasons why gliomas often have a bad prognosis is because of their high heterogeneity, which arises from differences affecting histological and molecular levels \cite{Nicholson2021}. Glioma samples are characterized by different molecular profiles and tumor cell types, which result in distinct glioma types. 
The criteria for glioma classification are constantly revised according to new discoveries. For many years, the diagnosis of glioma has been exclusively based on histology, being recently subjected to a substantial revision with the inclusion of genetic information in the classification procedure. Despite the importance of this continuous update, it represents a challenge for diagnosis and design of effective therapies \cite{Chen2017}.  

The World Health Organization (WHO) classification of the Central Nervous System (CNS) introduced molecular characteristics as part of the diagnosis of glioma tumors in 2016 \cite{WHO2016}. In particular, the status of the isozymes of the isocitrate dehydrogenase (IDH) gene family and the combined loss of the short arm chromosome 1 and the long arm of chromosome 19 (1p/19q codeletion) have been integrated with the former histological diagnosis. This novelty represented a valuable improvement, given the subjectivity of the histological evaluation. However, with this integrated diagnosis contradictory results could arise, in case of patients showing inconsistent histological and molecular features.

In 2021, the glioma classification procedure was updated \cite{WHO2021}, mainly focusing the diagnosis on the evaluation of objective genetic features, and solving the problem of possible contradictory diagnostic results that could arise from the 2016 approach. 
The IDH mutation status became central for determining the tumor type since samples exhibiting IDH mutation 
coupled with 1p/19q codeletion are classified as oligodendroglioma, while IDH-mutant samples not presenting 1p/19q codeletion 
are labeled as astrocytoma. IDH-wildtype samples 
could be classified as glioblastoma (GBM) if either certain histological features or genetic parameters are present. 
These modifications in the glioma classification as defined by the WHO CNS guidelines 
aimed to support as best as possible the diagnosis, which is the first essential step to designing an effective therapy \cite{Yang2022}. Indeed, given the deep differences in glioma, each type could be influenced by specific markers, having an impact on cancer development  \cite{Jones2021}. In this context, discovering the existing underlying characteristics 
could be crucial to improve patient prognosis \cite{Chen2017a}.

Nowadays, thanks to modern data extraction techniques and gene sequencing advances, large sets of data are available. The huge dimension of these datasets, as well as the hidden relations existing between biological entities, make it challenging to infer information linking variables and clinical outcome.
To this end, mathematical and statistical methods can be employed 
to tackle 
the problem and to identify potential therapeutic targets. 

The high dimensionality of the data poses challenging problems in model identification and parameter estimation, which can be addressed by means of regularization techniques. These methods consist in adding a penalty term to the cost function of the chosen statistical model. In recent years, regularization has been widely used in bioinformatics applications, and several methods have been developed to address this need \cite{Vinga2020}. 
The least absolute shrinkage and selection operator (lasso) \cite{Tibshirani1996} represents the most known regularization method. It considers as penalty the $\ell_1$-norm of the coefficients that are the solution of the regression problem. This forces some estimated coefficients to be zero, thus simultaneously performing feature selection and estimation. 
However, when the number of features ($p$) is much larger than the number of samples ($n$) the feature selection becomes arbitrary, e.g., among correlated features that can predict the outcome, lasso may choose to include one (or a subset) and exclude others randomly. 
Since the genes cannot be considered as independent features, as they are
part of interconnected networks, graph theory can effectively be employed
to determine the relations between them. As a results, network estimation
techniques have been widely used in precision oncology \cite{Zhang2017a, Pai2022}, also in the context of glioma \cite{Lopes2021}. 

Many techniques have been developed to estimate a graph starting from real data \cite{Qiao2018}.
In this work, we employed graphical lasso, a widely known method combining regularization and network estimation \cite{Friedman2008}. 
Graphical lasso estimates relations between variables introducing sparsity through lasso regularization. It is particularly accurate for huge datasets ($p\gg n $), and it has been applied in many fields \cite{Arbia2018, Krock2021,Menedez2010}.

Besides the identification of meaningful diagnostic features from the overall network, the evaluation of their prognostic role is an invaluable contribution. Regularized survival analysis can be used to this end. Statistical methods for survival analysis estimate the probability of the occurrence of a given event, e.g., death, in a certain period of time, taking into account patients' data, including clinical information \cite{Wang2019a}. Among them, we used the Cox regression model, which is able to determine the association of a set of features with the time until the event of interest occurs \cite{Cox1972}. 
Coupling survival analysis with regularization techniques allows for the selection of features carrying information about the overall network and survival and therefore promising as prognostic biomarkers.

In this work, we considered data from The Cancer Genome Atlas (TCGA) program, consisting of a huge dataset ($p \gg n$) of RNA-sequencing (RNASeq), from the LGG and GBM projects \cite{TCGA2015,Brennan2013,McLendon2008}. These datasets have been preprocessed to assess the graphical lasso hypothesis and to update the sample classification \cite{Silva2023} according to the 2016 and the 2021 WHO CNS guidelines. Through the graphical lasso method, we reduced the dimensionality of the dataset by selecting a subset of variables for each glioma type, and we analyzed the results by identifying key features (\emph{hubs}) within the estimated networks. We performed survival analysis through Cox regression model \cite{Cox1972} with regularisation with the lasso penalty \cite{Tibshirani1996}, by considering (i) the set of selected variables, (ii) the variables that have been exclusively selected from each glioma type, and (iii) the hub genes.
This approach allowed us to 
assess if the variables selected for their diagnostic relevance also carry prognostic information.

The paper is structured as follows. In the materials and methods section, we start introducing the basis of graphical lasso and Cox regression methods, and we describe the datasets and the workflow of our analysis. In the results section, the model outcomes are presented and analyzed. A comprehensive discussion links our results with the biological interpretation. Finally, the conclusion resumes our main contribution.

\section{Material and methods}\label{sec:M&M}

\subsection{Graphical lasso} \label{subsec:glasso}
Let $\mathbb{G}$ be a graph with $X=(X_1,...,X_p)$ nodes (variables) having a multivariate normal distribution $X \sim \mathcal{N}_p(0,\Sigma)$. Let $\Sigma$ and $S$ be the theoretical and empirical covariance matrix, respectively. 

The graphical lasso \cite{Friedman2008}, herein designated as glasso, finds the precision matrix $\Theta=\Sigma^{-1}$, by solving the following Gaussian log-likelihood maximization problem:
\begin{equation}\label{eq:glasso}
	\max_{\Theta} \{ \log(\det\Theta) - tr(S\Theta) -\rho ||\Theta||_1\},
\end{equation}
where $tr(\cdot)$ is the trace operator. The presence of the regularization term $\rho ||\Theta||_1$ serves as penalty, providing a weight on the elements of the matrix $\Theta$, and inducing sparsity into the solution. The value of the regularization parameter $\rho$ determines the degree of sparsity.
It has been proved that the network structure of the graph $\mathbb{G}$ can be estimated based on the non-zero entrances of $\Theta$ \cite{Lauritzen1996}. In particular, if $\Theta_{i,j}=0$, for certain $i$ and $j$ ($i\neq j$), it means that the variables $X_i$ and $X_j$ are conditionally independent given the others $X_k, \ k=1,...,n, \ k\neq i,j$.

\subsection{Cox regression modelling with lasso regularisation} \label{subsec:Cox}

Cox regression is a survival model that relates the rate of an event happening in a given time point (e.g., death) with several factors. 
Given $p$ features (e.g, gene expressions) for $n$ samples, and let $(Y_i, \ \delta_i)$, $i=1,...,n$ be, respectively, the survival time and the indicator of the occurrence of the event ($\delta_i \in \{ 0, \ 1 \}$)  of the \emph{i}-th patient. The goal of a survival models is estimating the hazard function, i.e., the failure rate depending on time. If we define that function as 
\begin{equation*}
    h(t|X^i)=h_0(t) \exp{X^i\beta},
\end{equation*}
where $h_o(t)$ is a baseline hazard function, 
the Cox regression model determines the coefficients $\beta=( \beta_1,...,\beta_p )$ that maximize the following partial log-likelihood function:
\begin{equation}
    \max_{\beta} \sum_{i=1}^n \delta_i \left[ X^i \beta - \log\left(  \sum_{j: Y_j \ge Y_i} X^j \beta\right)\right],
\end{equation}
where $X^i=\left(X_1^i,...,X_p^i\right)$ is the set of features of the \emph{i}-th patient. 

To study the relationship between predictor variables and survival outcome in a high-dimensional scenario ($p \gg n$), the Cox regression model can be regularised by including a penalty function $P(\beta)$. In this work, we considered lasso regularization \cite{Tibshirani1996}, therefore $P(\beta)=\lambda ||\beta||_1$.

\subsection{Dataset preprocessing and framework}\label{subsec:pipeline}
\begin{figure} [!ht]
	\centering
	\includegraphics[width=0.85\textwidth]{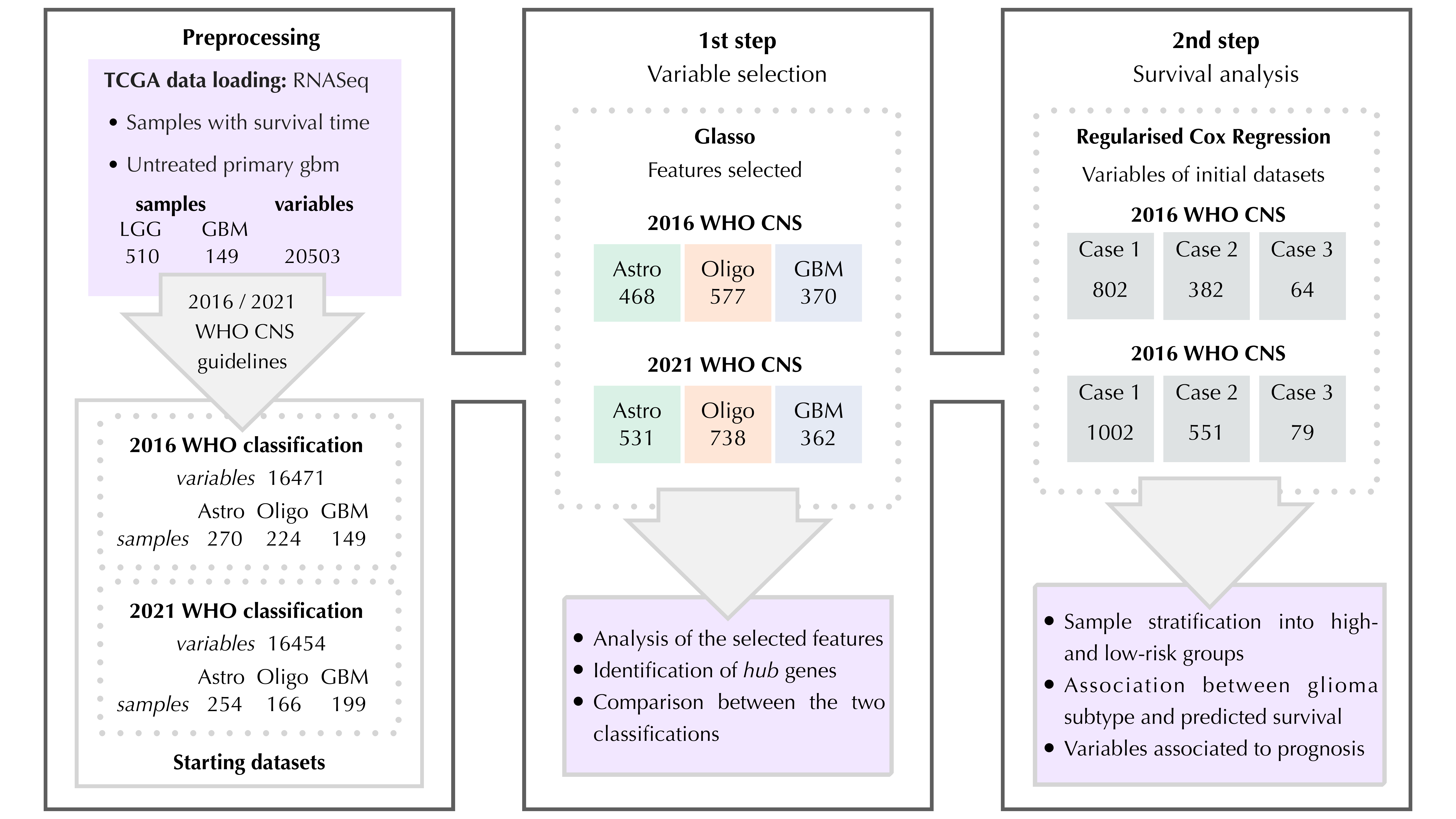}
\caption{Workflow of the analysis. After the preprocessing of the TCGA-LGG and TCGA-GBM databases, we obtained 6 datasets (3 for each classification procedure). (1st step) glasso algorithm has been applied separately on each glioma type, obtaining the network-based variable reduction. The number of features selected for any given type is reported in the second box of the pipeline. Green indicates astrocytoma (Astro), orange is oligodendroglioma (Oligo) and blue represents GBM. (2nd step) the features selected by the previous step are used as input for the survival analysis. All the datasets consider a matrix comprising all the glioma samples and different variables, depending on the case of study. \emph{Case 1} takes into account all the selected variables. \emph{Case 2} refers to the variables exclusively selected from each glioma type. \emph{Case 3} considers only the hub genes. Boxes in figure indicate the number of variables of the starting datasets. In each step, the analysis and the comparison between the two classification outcomes have been performed.}\label{fig:pipeline}
\end{figure}

The dataset selection procedure and the workflow of our analysis are summarized in Figure \ref{fig:pipeline}. The glioma dataset we used in this study is available on TCGA portal, under the LGG and GBM projects, and it has been downloaded by the \texttt{getFirehoseData} function of \textbf{RTCGAToolbox} package \cite{RTCGAToolbox2014}, from R software (version 3.5.1, https://www.r-project.org). 
This dataset contains expression levels from 20,503 genes, normalized by transcript per million and quartile normalizations. We updated the patient’ diagnosis according to both 2016 and 2021 WHO CNS guidelines, by integrating patient’ molecular profiles \cite{Ceccarelli2016} and reproducing the procedure explained in \cite{Silva2023}.
 We considered samples reporting the survival time information, in order to perform survival analysis. Regarding the GBM dataset, we considered samples related to ``untreated primary gbm".
 
To assess the assumption of normality required by glasso, we applied the normalization with \texttt{huge.npn} function from the \textbf{huge} R package \cite{huge2014}, which implements the gausianization through the nonparanormal transformation. For our study, we considered only the variables with normal distribution, according to the Jarque-Bera test \cite{Jarque2011}. Finally, following the 2016 WHO CNS classification, our reference dataset contains 16471 normally distributed variables for 270 astrocytoma, 224 oligodendroglioma, and 149 GBM samples. Following the 2021 WHO classification, the dataset is constituted of 16,454 normally distributed variables for 254 astrocytoma, 166 oligodendroglioma, and 199 GBM samples (Figure \ref{fig:pipeline}; Preprocessing).

For the first step of our methodology, glasso has been applied to each dataset, separately. The estimation of sparse networks led to a network-based variable selection, which has been mathematically validated in order to assess the reliability of the set of identified variables (see \ref{app:validation}).
The process of variable selection produced a different subset of genes for each glioma type, namely, 468, 577, and 370 genes for astrocytoma, oligodendroglioma, and GBM, considering 2016 WHO classification, whereas 531, 738, and 362 genes for astrocytoma, oligodendroglioma, and GBM, considering 2021 WHO classification (Figure \ref{fig:pipeline}; 1st step). These have been used as starting datasets to perform the second step of survival analysis, as explained in the following sections.
 
\subsection{Implementation of variable selection}
The variable selection was performed by applying glasso through the \texttt{huge.glasso} function of the \textbf{huge} R package \cite{huge2014}, by setting the regularization parameter $\rho=0.9$. This value has been empirically chosen in order to consistently reduce the number of variables and increase interpretability. 
Indeed, as mentioned in Section~\ref{subsec:glasso}, the regularization term determines the sparsity of the precision matrix $\Theta$.
As consequence, setting a high value of $\rho$, many variables will result unconnected within the graph, allowing for a network-based variable selection. To identify key nodes in each network, we introduced two different measures. 
 Given a variable $X_i$, the \emph{weight} measure consists in evaluating the degree of the \emph{i}-th node as:
\[m_i= \sum_{j=1, j\neq i }^n |\theta_{ij}|,\] 
while its \emph{count} measure represents the number of the connection of the variable: 
  \[m_i=\sum^n_{j=1, j\neq i} q_{ij},\]
  where $q_i=1$, if $|\theta_{ij}|>0$, $q_i=0$ otherwise.
As such, the genes can be sorted in ascending order according to both these measures, obtaining, for each dataset, two ranked lists of genes.

To identify the subset of the most relevant variables, for each list, we re-scaled the computed measures, by distributing them uniformly in $[0,\ 100]$. Then,
 let $p_i$ be the value of the measure $m_i$ converted in percentage, we defined $X_i$ as a \emph{hub} variable if $p_i>t$, for a certain threshold $t$. The threshold parameter choice is a crucial point of our analysis. 
Setting too low values could result in big sets of variables, which is hard to interpret.
 Conversely, too high thresholds could remove important variables, which will be omitted from our analysis. The dimension of the set of variables selected through glasso also influences this choice, since the larger the set is, the more we need to filter our results. 
We manually analyzed the glasso results, with the aim of identifying a discriminating value able to divide the list of variables into the hubs and non-hubs groups.
We observed that in astrocytoma and oligodendroglioma cases, the ranked list of genes reported very close percentage measures, i.e., the difference between the measures of two following variables was almost constant, 
leading to no obvious choice for the threshold. 
On the other hand, the variables selected from GBM dataset can clearly be divided into two groups, with a consistent variation of measure values around $60\%$. For this reason, we set $t=60$.
Each measure selects a subset of hub features, thus we considered as final hubs the union of these two subsets.

\subsection{Regularized Cox Regression Modelling, Patient Stratification and Survival Analysis}
Survival analysis has been performed by a regularized Cox regression model, implemented through the $\mathbf{glmnet}$ R package \cite{Friedman2010}, by considering 3 cases.
We considered a Pan-Glioma dataset, composed by all glioma samples (LGG+GBM) and (case 1) all the variables selected by glasso, (case 2) the variables exclusively selected by each glioma type, and (case 3) only the features identified as hubs. This results in 6 datasets, i.e., 3 cases for each glioma classification procedure (Figure \ref{fig:pipeline}-2nd step).

The parameter $\lambda$ of lasso penalization during model fitting was determined by $pmax$, a parameter limiting the maximum number of candidate features to be selected to avoid overfitting. For this study, we have set $pmax$ based on the 10 EPV (events per variable) rule of thumb \cite{Harrell1984, Harrell1996,Steyerberg2000}, which relates the number of events with the maximum number of predictors that can be studied. 
Therefore, $pmax$ was set as the ceiling function of the number of patients with the outcome event (death) divided by 10. 
The 2016 and 2021 WHO CNS classifications led to datasets composed by 233 and 226 events, which resulted in \emph{pmax}=24 and \emph{pmax}=23, respectively. Once defined $pmax$, the value of the parameter $\lambda$ has been computed through the function \texttt{glmnet}, which determined the appropriate regularization parameter leading to the chosen number of variables.
The dataset dimensions and the parameters chosen in each case are summarized in Table \ref{tab:SAparameters}.

The function \texttt{separate2GroupsCox} from the $\mathbf{glmSparseNet}$ R package \cite{Verissimo2018} 
was used to separate the data into two groups of patients: high- and low-risk. This function divides patients according to their 
prognostic index (PI), resulting from the matrix multiplication of the data with the fitted coefficients. By default, patients are ordered and stratified according to the median value of the PI distribution, i.e., a patient is attributed to the low-risk group if its corresponding PI (computed using Cox model) is below or equal to the median risk, and assigned to the high-risk group otherwise.
However, in all the defined datasets, the prognostic indexes exhibited a bivariate normal distribution, not centered by the median value (Figure \ref{fig:suppPI}). 
Consequently, considering the default stratification is not suitable for our study, since it could force samples to belong to the wrong group.

To determine the threshold value in this imbalanced class situation, we computed a kernel density estimation of the distribution of the PI by the \texttt{density} R function, and we extracted the local minimum of such distribution between the two gaussian peaks ($\widehat{PI}$). 
Samples with PI $\le \widehat{PI}$ were assigned to the low-risk group, while samples having PI $>\widehat{PI}$  were considered to be part of the high-risk group. Supplementary Figure \ref{fig:suppPI} shows the distribution of prognostic indexes and the corresponding threshold, in each case. 

Based on this patient' stratification, Kaplan-Meier survival curves were drawn and statistically compared with the log-rank test (hypothesis test to compare the distribution of time until the occurrence of an event in independent groups). A significance level of 0.05 was considered.

\section{Results}\label{sec:results}

In this section, the results obtained from variable selection and survival analysis are shown. We discuss the outcomes in light of the comparison between the 2016 and 2021 WHO CNS guidelines, highlighting differences and similarities arising from the transcriptomics data. Finally, we used the obtained results as baseline knowledge to explore the 2021 estimated networks.

\subsection{Variable selection} \label{subsec:variable-sel}
\begin{figure}[!ht]
	\centering
	\includegraphics[width=0.95\textwidth]{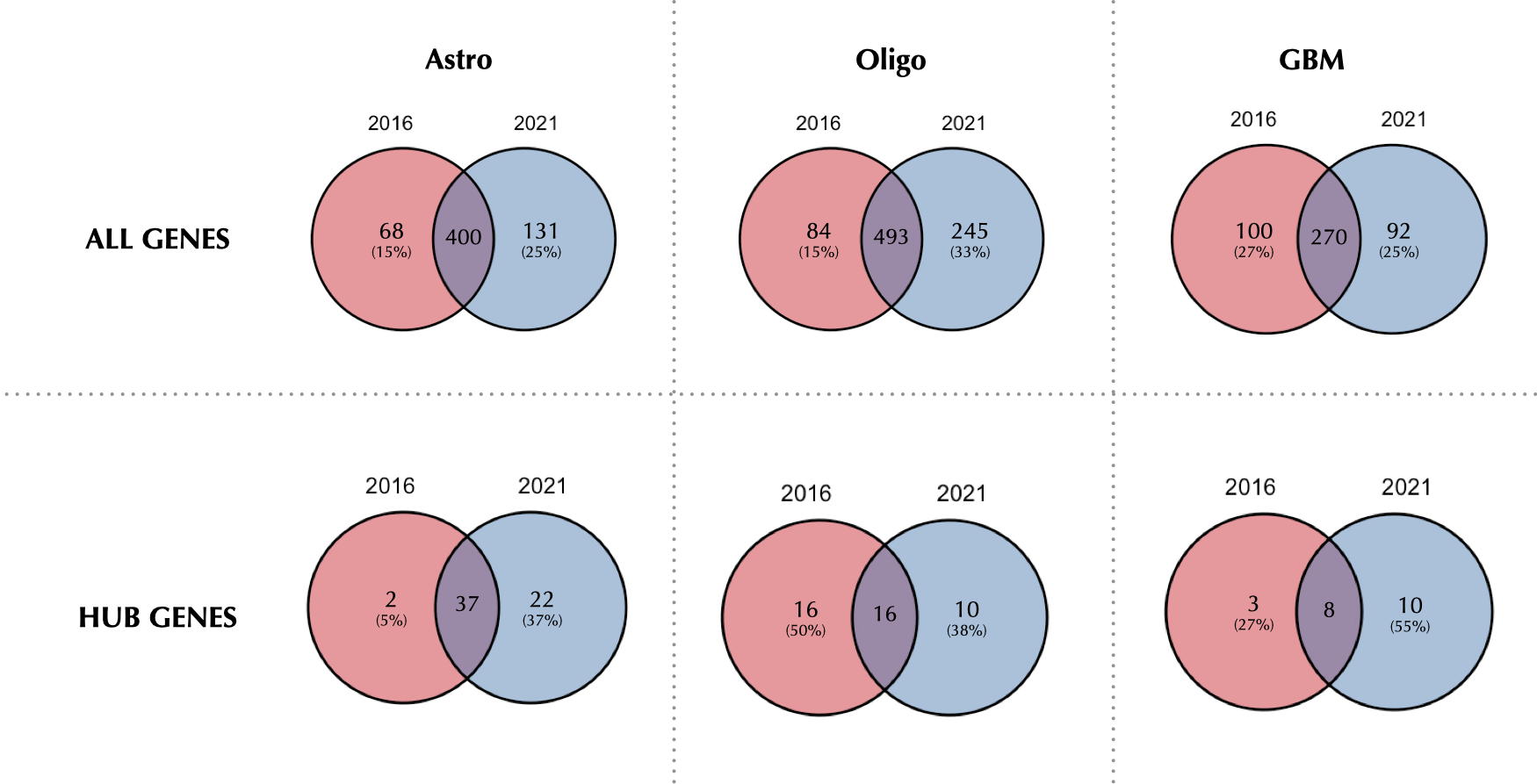}
	\caption{Venn diagrams showing the number of variables selected by glasso considering 2016 versus 2021 WHO CNS classification guidelines, for each glioma type (astrocytoma -- Astro --, oligodendroglioma -- Oligo --, and GBM).  Percentages in brackets refer to the relative percentage of the exclusively selected variables for each set. The first row considers all the selected genes, while the second row is only about hub genes.}\label{fig:so}
\end{figure}

We performed the variable selection by considering each glioma type, separately, taking into account two classification procedures. We compared the genes which have been selected for a given tumor type, classified according to the two different guidelines (Figure \ref{fig:so}; first row). For astrocytoma and oligodendroglioma, the number of variables selected by considering the 2021 WHO CNS classification was larger than the one obtained by following the 2016 guidelines. 
On the other hand, looking at the output derived from the 2016 classification, the percentage of variables exclusively selected for GBM is double the one of LGG (astrocytoma and oligodendroglioma). 
However, for all the types, the ranking of the genes that have been exclusively selected is very low, considering both weight and count measures, and, among these genes, there are no variables identified as hubs. 
The analysis has been repeated by considering the hub gene subsets to discuss the differences in selecting important features. Figure \ref{fig:so}; second row shows the result of this comparison. 
For astrocytoma and GBM, most 2016 hub genes are identified as important also according to the 2021 WHO CNS sample classification, \emph{i.e.}, $95\%$ and $73\%$ of astrocytoma and GBM 2016 hubs, respectively. In both cases, the 2021 classification led to a larger set of key features compared with 2016. 
 Differently, for oligodendroglioma, our procedure selects more hub genes by starting from the dataset obtained following the 2016 WHO CNS guidelines, with almost $50\%$ of key genes shared by both classifications.
Overall, all the hub genes (exclusive hubs included) are in the subset of selected variables of the corresponding glioma type, regardless of the classification procedure.

\begin{figure}[!ht]
	\centering
	\includegraphics[width=0.65\textwidth]{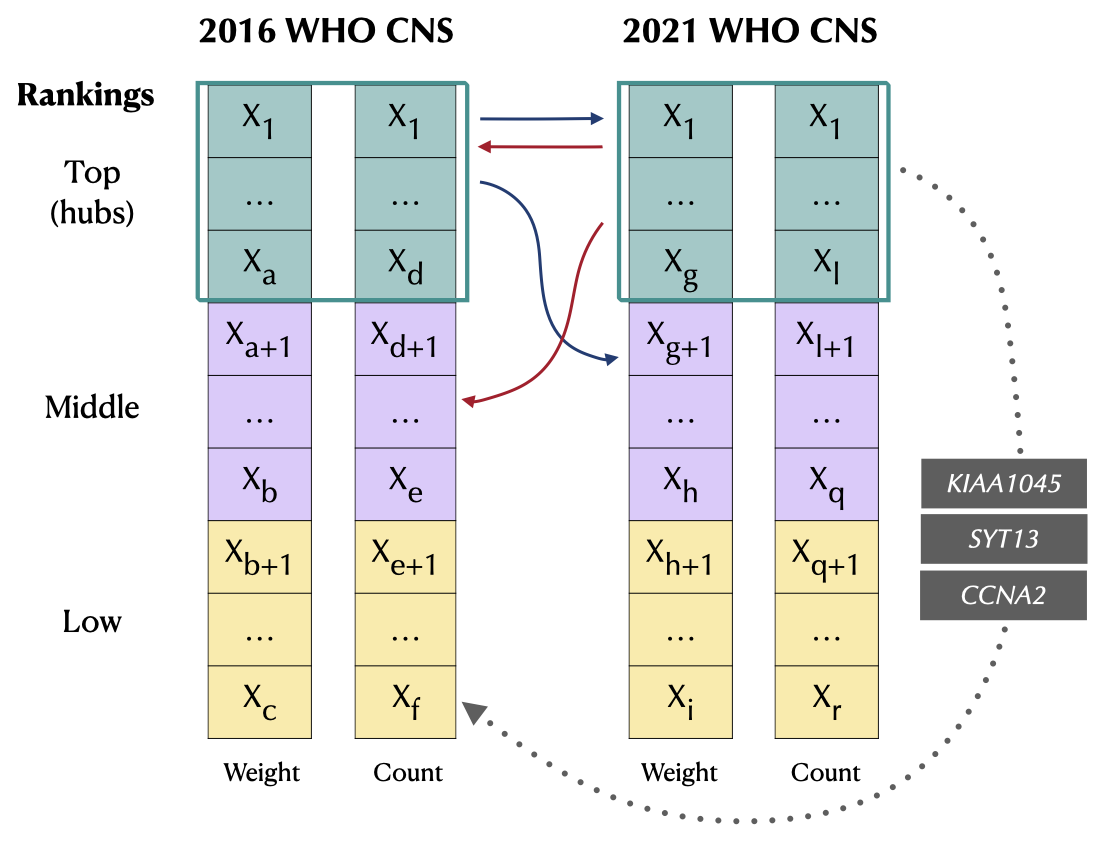}
	\caption{Cross-comparison of the hub gene rankings with respect to the other classification. Green, purple and yellow vector sections represent top, middle and low positions, respectively. Continuous arrows indicate the positions of most of the hubs. Dotted arrows highlight the three 2021-hub genes having low 2016-rankings. }\label{fig:so_comparison}
\end{figure}

Given the genes identified as hubs exclusively for a certain classification, we can check their rankings 
computed in the other classification. Commonly, these genes are placed in high positions (green and purple boxes in the vectors in Figure \ref{fig:so_comparison}), suggesting that the hubs have a key role in the networks, independently from the considered classification.
Only a few genes do not follow this trend (dotted arrows in Figure \ref{fig:so_comparison})
In particular, in 2021 GBM we identified \emph{CCNA2} as hub, which has an average percentage measure of $2$ in 2016 GBM. \emph{KIAA1045} and \emph{SYT13} have been recognized as hubs in 2021 astrocytoma, but the average of their 2016 percentage measures is 4 and 8, respectively.  
The research literature on these genes revealed that all of them have been associated with glioma by different bioinformatic analyses \cite{Wang2012, Dai2017}, and \emph{CCNA2} has been recently proposed as an immunological biomarker for GBM \cite{Jiang2022}.

\begin{figure}[!ht]
	\centering
	\includegraphics[width=0.9\textwidth]{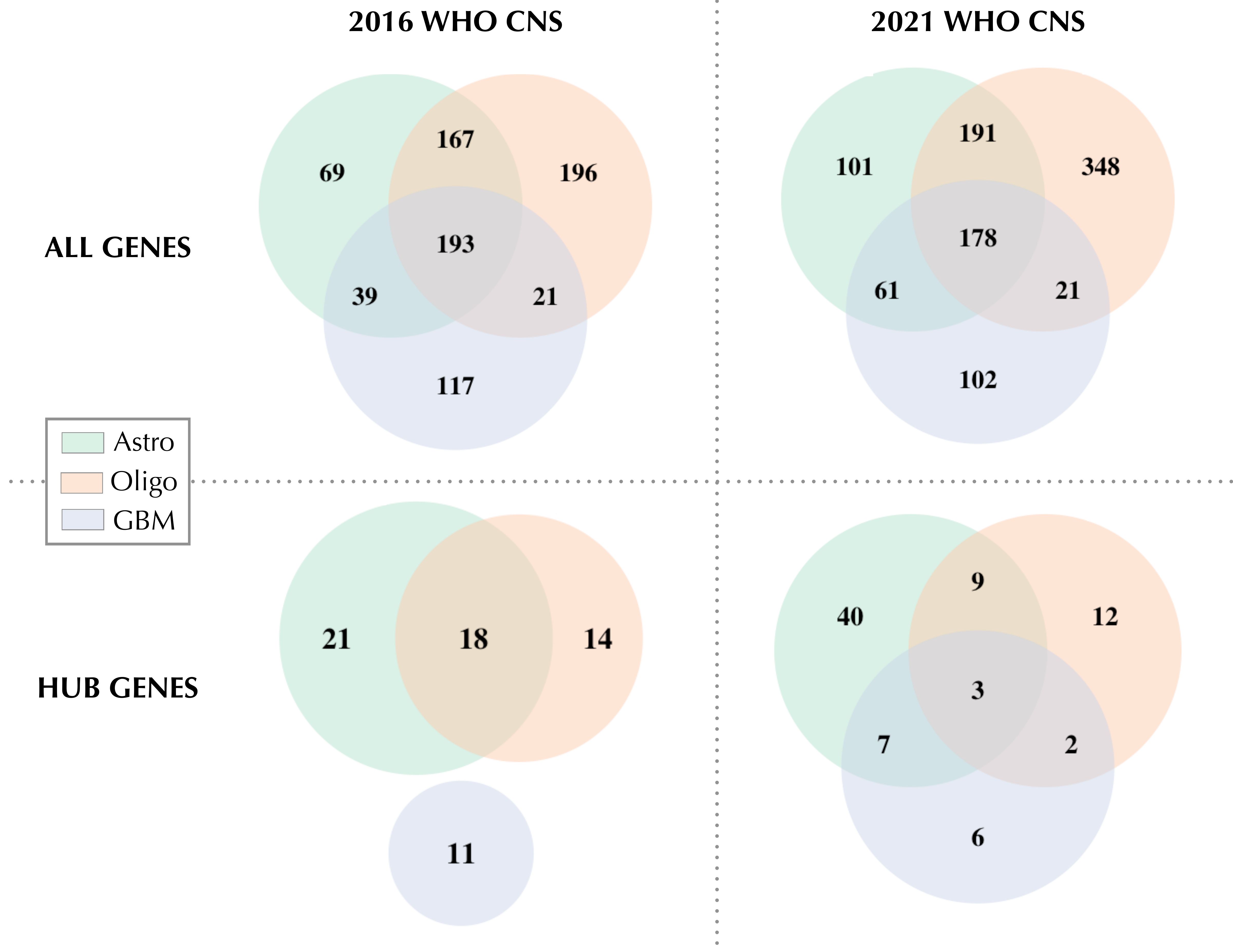}
	\caption{Venn diagrams comparing the variables selected by glasso from each glioma type (astrocytoma -- Astro --, oligodendroglioma -- Oligo --, and GBM). The first and second columns show glasso results respectively for 2016 and 2021 WHO CNS classifications. The first row considers all the selected genes, while the second row is only about hub genes.}\label{fig:RNASeq-results}
\end{figure}

 As shown in Figure \ref{fig:RNASeq-results} (first row), for both classification procedures, the three tumor types share a considerable number of features.
 Nevertheless, we can observe more similarities in gene selection between astrocytoma and oligodendroglioma, compared to GBM. This result has been also confirmed by the analysis of the hub genes (Figure \ref{fig:RNASeq-results}; second row), especially following the 2016 WHO CNS classification, where the set of GBM hubs does not intersect with the others.
The comparison of these diagrams also show a big difference in hub selection, in terms of shared/exclusive genes, revealing the impact of the classification procedure in the selection of key features.
 
\subsection{Survival analysis}
For each classification, survival analysis was performed by considering datasets comprising all the glioma samples and (case 1) the complete set of variables selected by glasso, (case 2) only the subset of variables exclusively selected from each glioma type, or (case 3) only the identified hub genes. We should note that the dataset construction depends on the class labels, as the variables constituting each dataset have been selected by applying glasso algorithm on each glioma type, separately. However, the Cox regression model fitting followed an unsupervised approach, i.e., the model learnt through data that was unlabelled.
Since the glioma types often exhibit different prognosis, as validation of our results, we expect to find an association between the given diagnostic label and the prediction of survival.

\begin{figure}[!ht]
	\centering
	\includegraphics[width=1.1\textwidth]{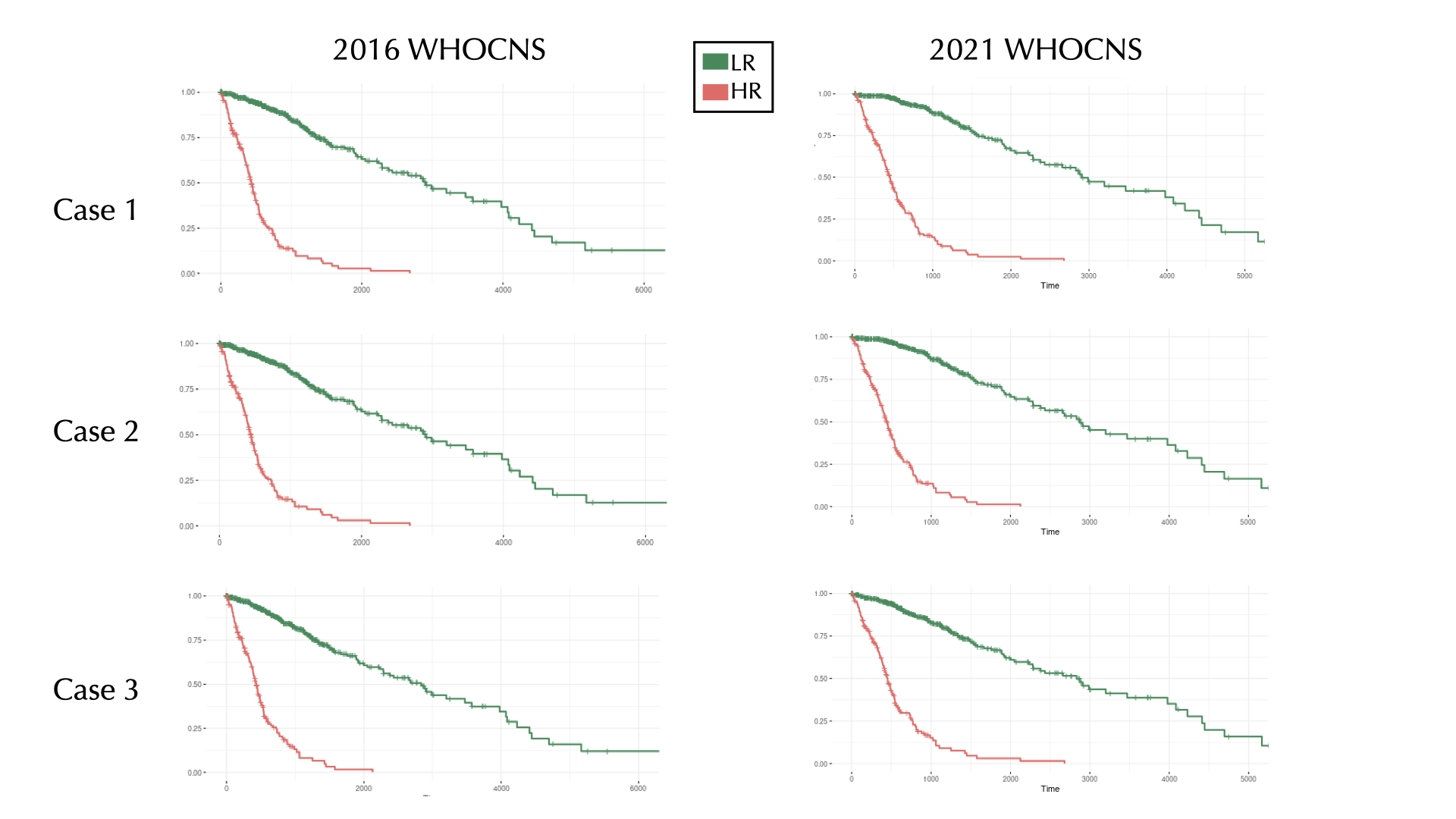}
	\caption{Kaplan-Meier survival curves obtained by the regularized Cox regression method applied on our cases of study. Case 1 considers a starting dataset comprising all the selected variables. Case 2 refers to the variables exclusively selected from each glioma type. Case 3 considers only the hub genes. Green lines represent the survival probability of low-risk (LR) sample group, while red lines refer to the high-risk (HR) sample group. }\label{fig:SAcurves}
\end{figure}

Overall, our sets of variables allow a statistically significant separation (\textit{p}-value $< 10^{-16}$) of the two high- and low-risk groups, as shown in the Kaplan-Meier survival curves in Figure \ref{fig:SAcurves}. We observe that, although case 2 considers a widely reduced dataset than case 1 (Table \ref{tab:SAparameters}), the survival results are comparable. Moreover, the 2021 WHO CNS classification leads to better results than 2016, since, in every case, the two curves have a clearer separation.  

\begin{figure}[!ht]
	\centering
	\includegraphics[width=\textwidth]{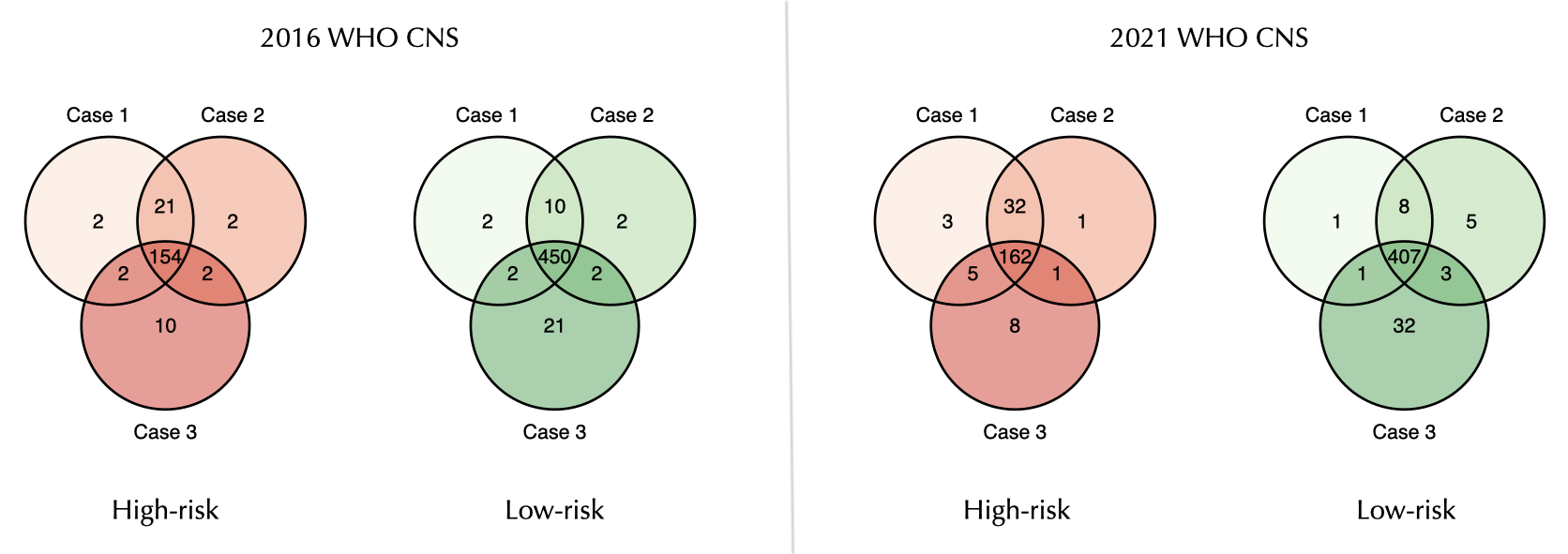}
	\caption{Venn diagrams comparing the composition of the high- (red) and low-risk (green) groups identified in each case, for a given glioma classification. 
 Case 1 considers a starting dataset comprising all the selected variables. Case 2 refers to the variables exclusively selected from each glioma type. Case 3 considers only the hub genes. }\label{fig:high-low}
\end{figure}

To analyze the two high- and low-risk groups, we first compared the patient stratification obtained in the three cases (Figure \ref{fig:high-low}). Overall, for both the classifications, most of the samples are recognized to be part of the same group,i.e., about $90\%$ and $78\%$ of samples are stably in the low- and high-risk groups, respectively, even by changing the reference dataset. As consequence, to explore the tumor types associated with the samples constituting these two groups, we can focus on one of these cases without loss of generality. In particular, we discuss case 1, as its starting dataset includes both the sets of variables of cases 2 and 3, increasing the reliability of the corresponding results.

Table \ref{tab:highlowrisk-samples} resumes the numbers and percentages of high- and low-risk samples identified in case 1, by specifying their glioma type. 
 Considering the 2016 variables, the high-risk group contains 179 samples. According to 2016 WHO CNS, $82.7\%$ of them ($n=148$) are classified as GBM, while the remaining $17.3\%$ are astrocytoma ($n=24$) and oligodendroglioma ($n=7$). Curiously, most of the LGG samples being part of this group, in 2021 change their classification to GBM.
 Conversely, the 2016-low-risk group is constituted of 464 samples, among which $99.8\%$ are LGG. Only one GBM sample is included in the 2016-low-risk group, which is diagnosed as astrocytoma in the 2021 classification. On the other side, the dataset obtained from the 2021 selected variables leads to a high-risk group of 202 samples. The patients classified as GBM constitute $93.6\%$ of them, while the other $6.4\%$ are astrocytoma. There are no oligodendroglioma samples in the 2021-high-risk group. The low-risk group is mainly constituted by LGG ($97.6\%$), but there are 10 GBM samples. If we look at the histological features reported by the TCGA clinical information, 5 over 13 2021-astrocytoma samples being part of the high-risk group were histologically evaluated as GBM, as well as 9 over 10 2021-GBM samples have histological features of LGG (5 astrocytoma, 2 oligodendroglioma and 2 mixed). 

\begin{table}
	\centering \footnotesize
	\begin{tabular}{ccccc}
 \cline{2-5}
& \multicolumn{2}{c}{\textbf{2016 WHO CNS}} & \multicolumn{2}{c}{\textbf{ 2021 WHO CNS}}\\[1ex]
 \hline
\textbf{Glioma} \ & \textbf{ High-risk} &\textbf{ Low-risk} & \textbf{ High-risk}&\textbf{ Low-risk}\\
\textbf{type} \ & \textbf{179 (28$\%$)}&\textbf{ 464 (72$\%$)} &\textbf{202 (33$\%$) }&\textbf{ 417 (67$\%$)}\\[1ex]
\hline
Astrocytoma \ & 24 (13$\%$) &246 (53$\%$) &13 (6$\%$) &241 (58$\%$)\\[1ex]
Oligodendroglioma \ & 7 (4$\%$) & 217 (47$\%$)&0 (0$\%$) &166 (40$\%$) \\[1ex]
GBM \ & 148 (83$\%$)& 1 ($<1\%$) &189 (94$\%$)&10 (2$\%$) \\
	\hline
	\end{tabular}
\caption{Cross-comparison between the composition of high- and low-risk groups and sample glioma types, depending on WHO CNS classification.}
\label{tab:highlowrisk-samples}
\end{table}

The regularised Cox regression model allowed for the selection of relevant features to describe the patient's survival data.
In case 1, 24 and 23 genes are identified, respectively from the 2016 and 2021 classification datasets. We observed that $50\%$ of the 2016 genes were also selected either in case 2 ($n=9$) or case 3 ($n=3$).
This percentage increases by considering the 2021 variables, reaching $74\%$ (15 genes in case 2, and 2 genes in case 3).
To investigate the role of these genes in glioma disease, we performed a literature review. 
Among the 35 genes, 12 are already known to have a role in glioma, while 13 are either linked to other cancer processes or pointed out in relation to glioma by bioinformatic studies. We did not find any cancer-related information for the remaining 10 genes. More details are summarized in Table \ref{tab:literature}. 

\subsection{Gene networks from 2021 WHO CNS classification}

The previous analyses denoted the 2021 WHO CNS classification as the most suitable for studying the glioma types. Moreover, survival analysis recognized the set of exclusive genes as very informative for prognosis, since, in 2021, they represent $74\%$ of the variables identified as important to predict the patient's survival. For this reason, in this section, we focus on the networks estimated through glasso on the 2021 dataset, 
to further explore the relations linking genes that exclusively characterize the different types. 

\begin{figure} [!ht]
	\centering
\includegraphics[width=0.9\textwidth]{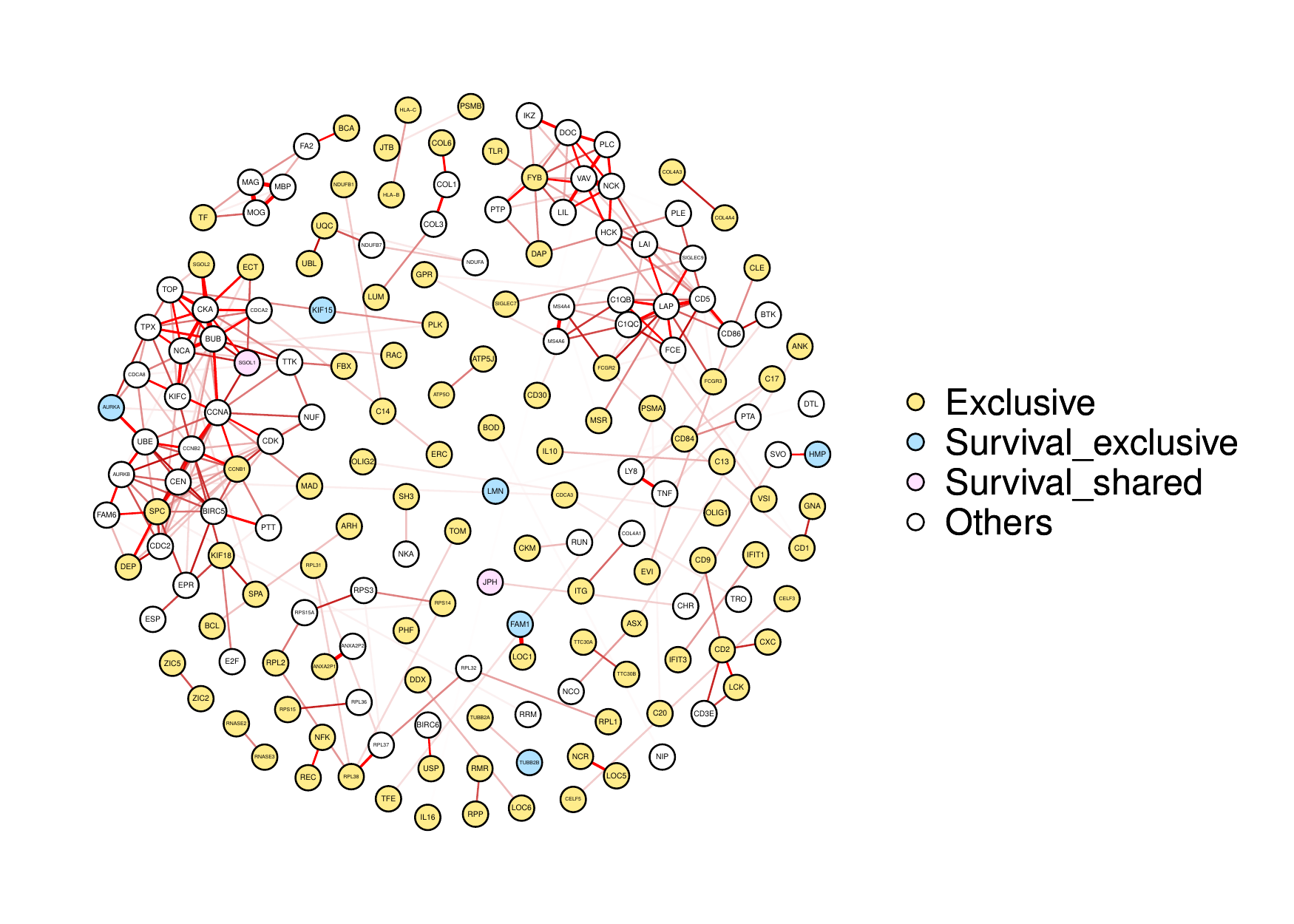}
\caption{GBM gene-network estimated through glasso. Nodes represent genes exclusively selected from GBM dataset (yellow) and others directly linked with them (white). Blue and pink nodes highlight genes with prognostic value, according to regularized Cox model, which could in turn be exclusive for GBM or shared, respectively. Network layout stresses the adjacency between nodes, based on the Fruchterman-Reingold force-directed algorithm. Edge thickness depends on the strength of the corresponding connection.  }\label{fig:gbmnet}
\end{figure}

\begin{figure} [!ht]
	\centering
	\includegraphics[width=\textwidth]{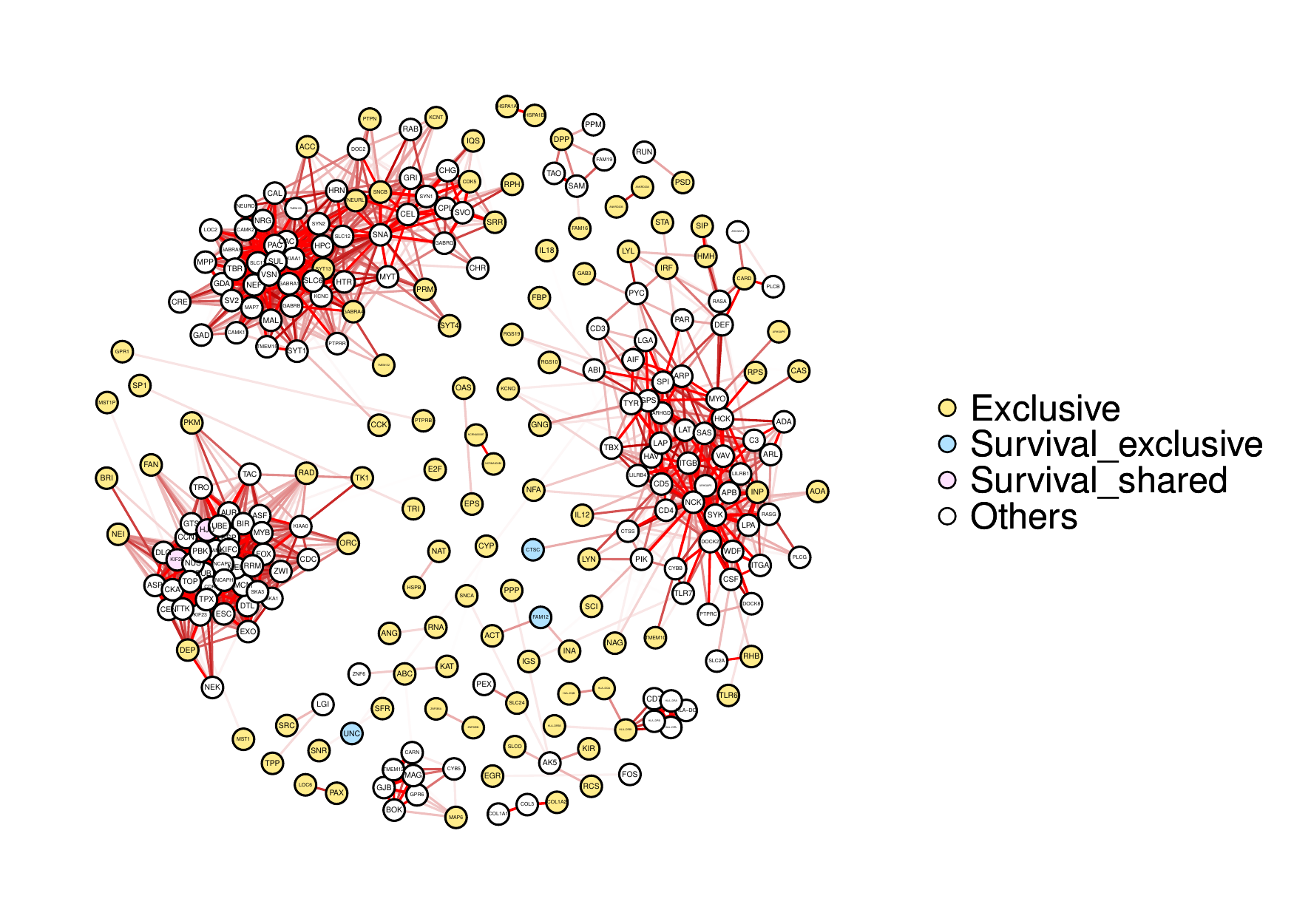}
\caption{ Astrocytoma gene-network estimated through glasso. Nodes represent genes exclusively selected from astrocytoma dataset (yellow) and others directly linked with them (white). Blue and pink nodes highlight genes with prognostic value, according to regularized Cox model, which could in turn be exclusive for astrocytoma or shared, respectively. Network layout stresses the adjacency between nodes, based on the Fruchterman-Reingold force-directed algorithm. Edge thickness depends on the strength of the corresponding connection.  
}\label{fig:astronet}
\end{figure}

\begin{figure} [!ht]
	\centering
	\includegraphics[width=\textwidth]{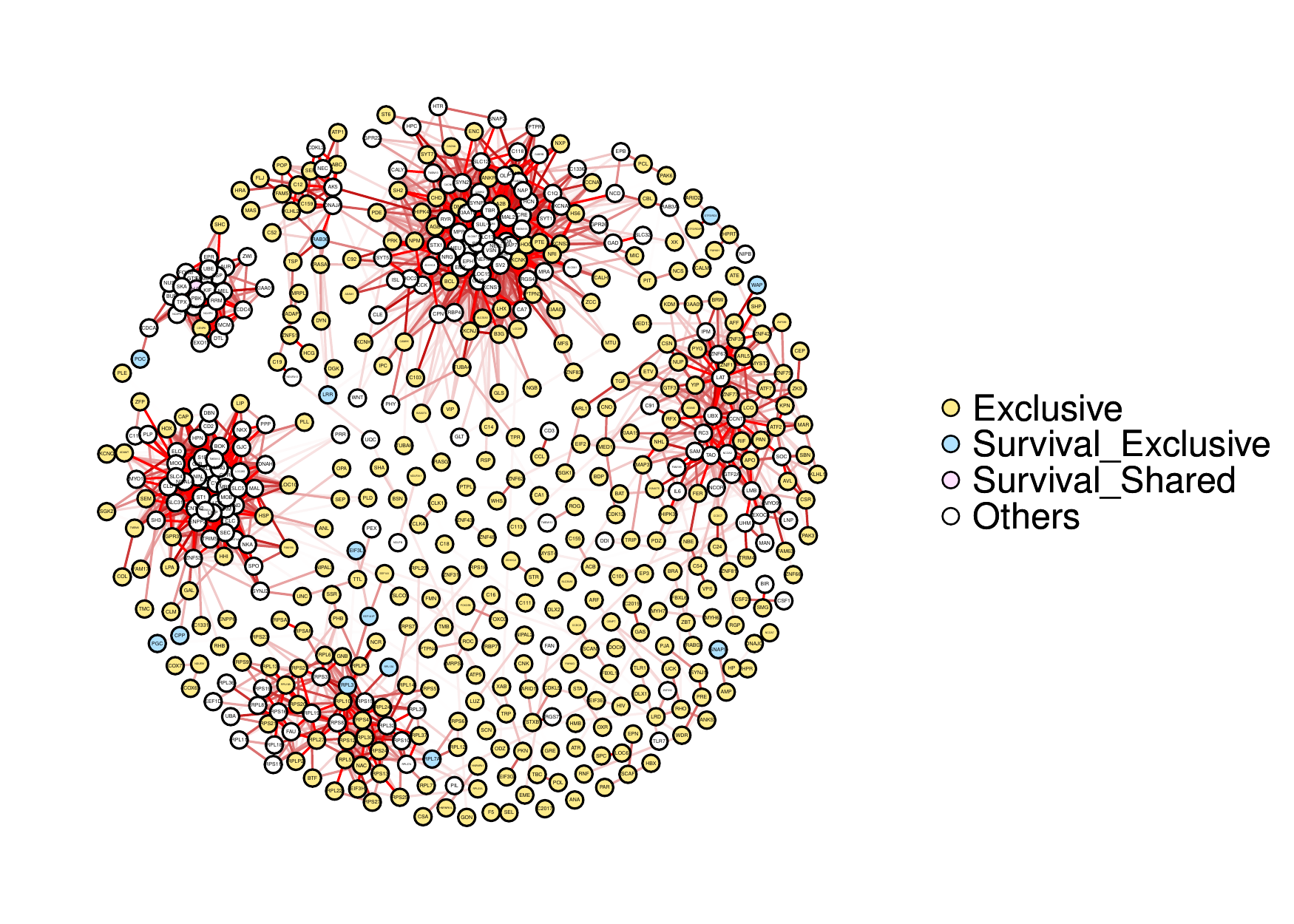}
\caption{Oligodendroglioma gene-network estimated through glasso. Nodes represent genes exclusively selected from oligodendroglioma dataset (yellow) and others directly linked with them (white). Blue and pink nodes highlight genes with prognostic value, according to regularized Cox model, which could in turn, be exclusive for oligodendroglioma or shared, respectively. Network layout stresses the adjacency between nodes, based on the Fruchterman-Reingold force-directed algorithm. Edge thickness depends on the strength of the corresponding connection. }\label{fig:oligonet}
\end{figure}

Networks in Figures \ref{fig:gbmnet}, \ref{fig:astronet}, and \ref{fig:oligonet} considered a subset of nodes composed of the genes exclusively selected for each glioma type (yellow), and the ones directly linked with them (white). Blue and pink nodes represent the genes selected as relevant in the survival analysis, which could be exclusive or shared by the tumor types, respectively.  
The chosen layout for network representation is based on the Fruchterman-Reingold force-directed algorithm implemented in \textbf{qgraph} R package \cite{qgraph-pack}, which distributes nodes based on the distance between them: adjacent nodes are close, while not-connected nodes are apart.

The GBM-network (Figure \ref{fig:gbmnet}) comprises 175 nodes, among which 102 are exclusive genes. On the left side, we can observe a dense group of nodes, including genes selected for their prognostic value (\emph{AURKA} and \emph{SGOL1}). The network includes 6 exclusive genes selected by the regularized Cox regression model, and some of them are involved in very strong connections, represented by bold edges. Specifically, \emph{HMP19} and \emph{FAM115C} are predicted to be linked with \emph{SVOP} and \emph{LOC154761}, respectively.
A strong correlation can be also observed between \emph{AURKA} and \emph{UBE2C}, which are two genes previously investigated for their combined effect in colorectal cancer \cite{Hedge2015, Ding2020}.

The astrocytoma-network has 262 nodes, but only 101 are exclusive of this tumor type (Figure \ref{fig:astronet}). 
Three big subnetworks can be identified in the graph. The one in left-bottom includes also genes selected in the regularized survival model, which are shared between the three glioma types (\emph{HJURP} and \emph{KIF20A}). 
Among the exclusive genes selected by regularized survival model (blue nodes), only 3 have been identified for astrocytoma. 
All the subnetworks are characterized by very strong connections, but they mainly involve shared genes.
A potential interesting small subnetwork is the one involving three exclusive genes, \emph{FAM123C} (selected in the regularized survival model), \emph{ACTL6B} and \emph{INA}. Interestingly, a study about MDA-9/Syntenin dysregulation in cancer pointed out these three genes as highly  downregulated in glioma with high MDA-9/syntenin expression \cite{Bacolod2015}.

The oligodendroglioma-network is the biggest one, with 552 nodes, and many exclusive genes ($n = 348$). We can observe 5 dense subnetworks, all of them including genes with prognostic value (blue or pink). This network contains most of the exclusive genes selected in the regularized survival model ($n = 13$). Interestingly, all the exclusive genes that in our literature review ended with no information are part of this network. One of them, \emph{RAB36}, appears strongly connected in biggest subnetwork with \emph{TSPYL5}, \emph{RASAL1},   \emph{KLHL26} and \emph{DNAJC6}, which are all exclusive except the last one.

\section{Discussion}
The main goal of this work was to discuss differences and similarities at the transcriptomic level between the two most recent glioma classifications, provided by the 2016 and 2021 WHO CNS guidelines.
The glioma type of each sample of TCGA RNASeq dataset were updated, and the glasso algorithm was applied to each glioma subgroup. The sets of selected variables were analyzed and used as the starting point for survival analysis performed with regularized Cox model and consequent patient stratification.

Our results suggest that the two classification procedures do not provide remarkable differences in variable selection.
Even if the algorithm identified variables exclusively related to one classification, the corresponding rankings are low.
Moreover, all the identified hubs are part of the subset of selected features, independently from the classification. 
However, comparing the hub subsets, our results highlight that the two procedures identified different key genes, suggesting that the classification could affect the detection of potential biomarkers specific to each glioma type. In particular, for GBM and astrocytoma, the 2021 classification provides larger sets of hub genes, including almost all the 2016 hubs. In addition, our analysis reveals three 2021 hub genes, one from GBM and two from astrocytoma, having low rankings according to the 2016 classification. This seems indicate that grouping the samples following the 2021 guidelines leads to a more comprehensive list of key features.

The statistically significant stratification of patients into high- and low-risk groups derived by survival analysis indicate that our selected sets of variables carry enough information to divide samples into high- and low-risk groups, serving also as validation of the glasso results.
Moreover, most of the variables selected as genes able to describe survival in case 1 (which took into account all the variables selected by glasso) have also been pointed out as output in case 2, in which the dataset considered only variables exclusively selected by each glioma type. This outcome is particularly interesting, since while the genes commonly selected by all the glioma types could be either related to specific-glioma mechanisms or general cell functions, the exclusive genes can be considered more characteristic of the corresponding glioma type. In this light, this results suggest that variables that have been selected based on their relevance for each glioma type (diagnostic value), they also hold prognostic information.

The patient stratification into high- and low-risk group derived from survival model fitting is consistent with literature \cite{Brown2022}, since GBM samples are associated with the worst survival, and they constitute the major part of the high-risk group. The 2021 classification leads to better results in terms of both distinction with statistical significance of low- and high-risk groups, and the association between GBM and poor survival, given that the percentage of GBM samples in the high-risk group increases by more than $10\%$ compared with 2016 sample grouping. 

As expected, the unsupervised sample grouping into high- and low-risk underlines the presence of some unconventional associations between glioma subtypes and overall survival. Based on the 2016 variable selection, the survival returns a high-risk group containing 179 samples. Despite $83\%$ of them being classified as GBM, few LGG samples are predicted to have poor survival. 
If we classify them according to the new 2021 WHO CNS guidelines, we observe that $84\%$ of these samples are diagnosed as GBM. Similarly, the only GBM sample being part of the 2016-low-risk group presents IDH-mutation, but no 1p/19q codeletion, which turns out to be astrocytoma in 2021. Although this result seems to support the idea that the 2021 classification leads to more consistent survival outcomes, this does not occur. Indeed, even if the number of LGG in the high-risk group is lower than the one predicted by considering the 2016 classification, there are 10 GBM assigned to the low-risk group. Interestingly, we can assess that 9 over 10 2021-GBM samples have histological features of LGG. 
This result is coherent with other studies that highlight the possibility of glioma samples presenting IDH mutation (which should be LGG in 2021) associated with poor outcomes, as well as IDH-wildtype (necessary condition to be GBM) with favorable survival \cite{Ceccarelli2016, TCGA2015}. 
All these results suggest that there are still unknown factors influencing survival, which could also affect tumor histology.

The literature research on the 37 genes identified as having prognostic relevance revealed that $34\%$ of them are known to have a role in glioma progression. Other $20\%$ have been identified as key in other cancers, while $17\%$ have been recognized as important by in silico analyses on glioma datasets. The remaining $29\%$ features are not linked to cancer-related processes.
The genes not previously investigated in glioma could be suitable for further investigation. Moreover, the fact that most of the genes identified in this study either arose from bioinformatic studies on glioma or are involved in cancer processes supports the necessity of validating them biologically.
Furthermore, network visualization could be a functional tool to disclose unknown relations and support biological research, allowing the detection of relevant subnetworks within gliomas types, and fostering validation studies.

\section{Conclusion}
To our knowledge, this is the first work discussing differences and similarities between the two most recent glioma classifications in terms of their impact on potential biomarker identification for each glioma type.

According to our results, transcriptomics data seems to align with the current changes in glioma classification guidelines, since the new 2021 WHO CNS classification provides better results in terms of variable selection and identification of key features, compared with the previous version of 2016.  

The survival analysis allows the separation of samples into the high- and low-risk group, proving that the set of variables we identified carry information also from the prognostic point of view. The 2021 classification led to better stratification of patients based on their survival profiles, since the percentage of GBM (characterized for worst prognosis than LGG) in the high-risk group increased compared with the 2016 classification, as expected. 
Moreover, with our pipeline, we provide a useful way to disclose unknown gene relations though network visualization.

Overall, our study brings insight into new features with a diagnostic and prognostic value that can be further biologically evaluated.
However, the presence of unexpected associations between glioma types and predicted survival suggests that, although molecular features are essential in predicting patient survival, histology has an impact on the risk of death. Therefore, additional efforts are needed to further characterize the heterogeneity of glioma types and improve the classification procedure. 

 We hope our results motivate the scientific community to further investigate the role of the genes we identified, intending to improve glioma therapies.

\section*{Author contributions}
\noindent \textbf{Roberta Coletti:} Conceptualization, Methodology, Software, Validation, Formal analysis, Writing - Original Draft, Visualization. \textbf{Mónica L. Mendonça:} Methodology, Software, Validation, Writing - Review and Editing. \textbf{Susana Vinga:} Conceptualization, Methodology, Supervision, Writing -  Review and Editing. \textbf{Marta B. Lopes:} Conceptualization, Methodology, Supervision, Writing -  Review and Editing, Funding acquisition. 

\section*{Acknowledgements}
This work was supported by national funds through Fundação para a Ciência e a Tecnologia (FCT) with references $CEECINST/00102/2018$, $UIDB/00297/2020$ and $UIDP/00297/2020$ (NOVA MATH, Center for Mathematics and Applications), $UIDB/50021/2020$ (INESC-ID), $UIDB/04516/2020$ (NOVA LINCS), and the research project “MONET – Multi-omic networks in gliomas” ($PTDC/CCI-BIO/4180/2020$).
\bibliographystyle{ieeetr} 
\bibliography{arxiv_template.bib}
\newpage
 \appendix
 \setcounter{table}{0}
\setcounter{figure}{0}

 \section{Mathematical validation}\label{app:validation}
 \vspace{-0.15cm}
 Mathematical validation of variable selection derived from glasso results has been performed to appraise the consistency of the solution of the equation \eqref{eq:glasso}. Let $D$ be the RNASeq dataset constituted by the subset of selected variables for a given glioma type. Setting a very low value of the regularization parameter $\rho$, we can determine a matrix $\Theta_D$ that does not lead to further variable selection\footnote{The value of $\rho$ should be different for any glioma-type, since everyone led to a different set of variable selected, determining different dimension of the dataset $D$.}. 
 Considering the objective function as a difference between two functions of $\Theta$, i.e. $F(\Theta)-R(\Theta)$, where $F(\Theta)=\log(\det\Theta) - tr(S\Theta)$, and $R(\Theta) = \rho ||\Theta||_1$, we compared the values of $F$ obtained from the dataset $D$ ($F_D$), with the one computed from 1000 random variable selections of the same dimension ($F_R$). If $F_D>F_R$ for all 1000 random datasets, we can conclude that the set of selected variables really led to the optimal solution. The results of the mathematical validation performed for both glioma classifications and for each glioma type are reported in Table \ref{tab:validation}, together with the fixed regularization parameters. 

We noticed that, although the optimization results depend on the regularization term $R(\Theta)$, choosing low values of $\rho$ we obtain $ O(F(\Theta)) \gg O(R(\Theta)) $, such that the regularization term can be ignored.
\begin{table}[h!]
\footnotesize
\centering
\begin{tabular}{ c |c|c ccc|  c ccc }
\cline{2-10} 
& \multicolumn{9}{c}{\textbf{2016 WHO CNS}} \\
\cline{2-10} 
\multicolumn{1}{c|}{}  &\multirow{2}{*}{$\rho$}&  \multirow{2}{*}{$F_D$ }    &   \multicolumn{3}{c}{$F_R$} &    \multirow{2}{*}{$R_D$ }    &   \multicolumn{3}{c}{$R_R$}  \\

\multicolumn{1}{c|}{} & &  &  Best & Average & Median  &  &  Best & Average & Median  \\
\hline
Astro &0.05& 476.12 &  76.69 & 52.93   & 53.24 & 0.79 & 0.62 &  0.64 &  0.63\\
  Oligo & 0.08 & 496.62  &  76.58 & 50.65 & 50.69 & 0.85 &0.74 & 0.73 &0.73\\
 GBM &0.02& 490.21 &   239.8590 & 225.44  & 225.44 & 0.68 &0.57 &0.57&0.57\\

\hline
& \multicolumn{9}{c}{\textbf{2021 WHO CNS}} \\
\hline
 Astro & 0.08 &  453.46 & 10.68 & -9.69 &  -9.83  &0.86 &0.74& 0.73& 0.73\\
  Oligo & 0.13 & 467.82  &  -6.27 & -32.43 & -32.17 &0.90&0.86 &0.83 &0.83\\
 GBM & 0.05 & 347.33 & 51.09 &   34.36 &  34.15 &0.75 &0.61 &0.61&0.61\\
\hline
\end{tabular}
\caption{Mathematical validation results. For each WHO CNS classification, the values of $F(\Theta)$ and $R(\Theta)$ functions are reported. $F_D$ and $R_D$ are computed by considering the dataset with the selected variables only. $F_R$ and $R_R$ summarize the results obtained for random variable subsets by showing the best, the average and median values computed. Table also shows in the first column the values of the regularization parameter $\rho$ fixed in each case. Astro: astrocytoma; Oligo: oligodendroglioma; GBM: glioblastoma.
}\label{tab:validation}
\end{table}

 \section{Additional figures and tables}  \label{app:tables}

 \vspace{-0.3cm}
 \begin{figure}[!ht]
	\centering
	\includegraphics[width=\textwidth]{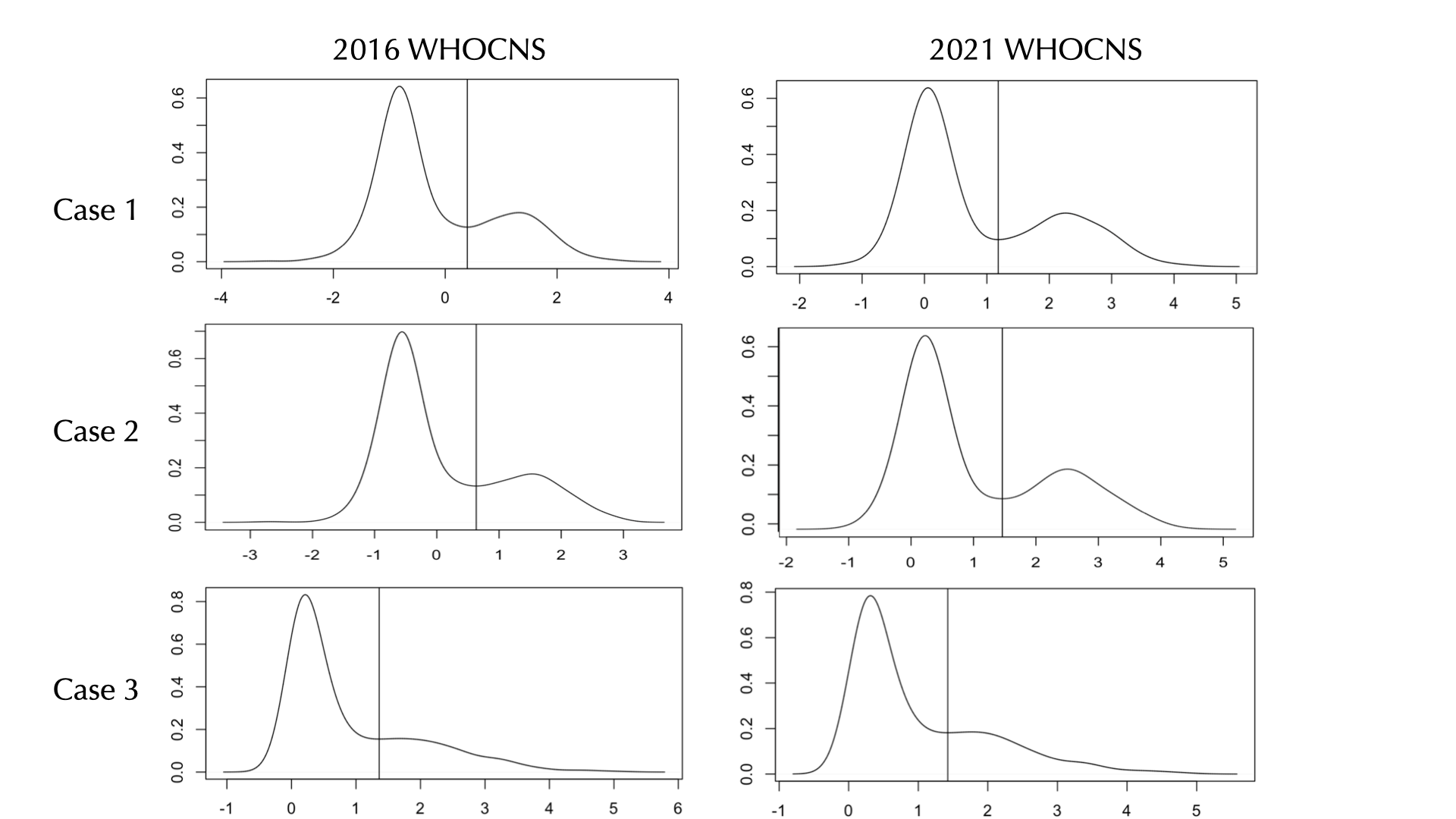}
	\caption{Density functions of the prognostic indexes (PI) of the samples constituting the dataset in each case.   
Case 1 considers a starting dataset comprising all the selected variables. Case 2 refers to the variables exclusively selected from each glioma subtype. Case 3 considers only the hub genes. 
 The vertical lines represent the chosen thresholds $\hat{PI}$ to divide samples into high- and low-risk groups. Samples with PI $\le \hat{PI}$ were assigned to the low-risk group, conversely, they are assigned to the high-risk group. 
  }\label{fig:suppPI}
\end{figure}

\begin{table}[h!]
	\centering \footnotesize
	\begin{tabular}{c|  c c c}
	\hline
 \textbf{WHO CNS} & \textbf{Case} & \textbf{Dataset dimension }  &\textbf{$\lambda$} \\
 &&($n$ x $p$)&\\
 \hline 
& 1 & 643 x 802 &  0.062\\[1ex]
2016 & 2 & 643 x 382  & 0.059\\[1ex]
&  3 & 643 x 64 & 0.010\\[1ex]
\hline
  & 1 & 619 x 1002 &  0.059\\[1ex]
2021 & 2 & 619 x 551 &  0.054\\ [1ex]
&  3 & 619 x 79& 0.014\\
	\hline
	\end{tabular}
\caption{For each case, dataset dimension and parameters setting are reported. Case 1 considers a starting dataset comprising all the selected variables. Case 2 refers to the variables exclusively selected from each glioma subtype. Case 3 considers only the hub genes. \emph{pmax} has been computed by dividing the number of dead samples by 10 (EPV). $\lambda$ is the Lasso regularization parameter.}
\label{tab:SAparameters}
\end{table}

	\begin{longtable}{c c m{6.5cm} c}
		\hline
		\footnotesize	\textbf{Gene} &  \footnotesize\textbf{WHO CNS} &\footnotesize \centering\textbf{Knowledge}  & \small \textbf{Ref.} \\ 
	\hline
		
		\small  \emph{AGAP4} & \small 2016 &\small \centering -- & \small -- \\ 
  \hline
    \small  \emph{HMP19} & \small 2016 &\small  \centering --  & \small --\\
   \hline
     \small  \emph{USP34} & \small 2016 &\small Downregulation inhibits pancreatic cancer growth and migration    & \small \cite{Lin2020}\\
   \hline
  \small  \emph{SAMD8} & \small 2016 &\small \centering --   & \small -- \\
   \hline
   \small  \emph{TMEM176A} & \small 2016 &\small  Promoter of GBM cell growth. & \small \cite{Liu2018}\\ 
  \hline
     \small  \emph{MYBL2} & \small 2016 &\small Overexpression induces esophageal squamous-cell carcinoma proliferation & \small \cite{Qin2019}\\
   \hline
     \small  \emph{OSCAR} & \small 2016 &\small Bioinformatic analysis reveled the expression of this gene correlates with poor prognosis in multiple cancer (including glioma)  & \small \cite{Liao2021} \\
   \hline
  \small  \emph{RRM2} & \small 2016 &\small It contributes to the migration and proliferation of glioma  & \small \cite{Sun2019}\\
   \hline
    \small  \emph{CLEC18B} & \small 2016 &\small  Overexpression induces proliferation, migration, and bad prognosis of GBM & \small \cite{Guo2018}\\
   \hline
     \small  \emph{PABPC3} & \small 2016 &\small  Possible gene associated with predisposition to breast cancer in North African population & \small \cite{Hamdi2018}\\
   \hline
  \small  \emph{CDC6} & \small 2016 &\small Upregulated in glioma. Correlated with immune infiltrates and poor survival. & \small \cite{Zhao2021,Wang2022} \\
   \hline
     \small  \emph{CLEC18A} & \small 2016 &\small \centering --   & \small -- \\
   \hline
   \small  \emph{WAPAL} & \small 2021 &\small \centering --   & \small --\\ 
   \hline
   \small  \emph{RPL13A} & \small 2021 &\small  Bioinformatic analysis revealed it is deferentially expressed in glioma. & \small \cite{Kreth2010}\\
   \hline
      \small  \emph{JPH3} & \small 2021 &\small  Tumor suppressor in hepatocellular carcinoma. & \small \cite{Huang2022}\\ 
   \hline
      \small  \emph{TUBB2B} & \small 2021 &\small Related to GBM invasion  & \small \cite{Yeini2021} \\ 
   \hline
      \small  \emph{HJURP} & \small 2021 &\small Involved in GBM proliferation and radioresistance. Terapeutic target.  & \small \cite{Serafim2020}\\ 
   \hline
      \small  \emph{CPPED1} & \small 2021 &\small Prevent bladder cancer progression.  & \small \cite{Zhuo2013}\\ 
   \hline
      \small  \emph{ARHGAP11A} & \small 2021 &\small  High expression correlates with better prognosis in gastric cancer. It determines malignant progression of hepatocellular carcinoma & \small \cite{Fan2021,Dai2018}\\ 
   \hline
      \small  \emph{RAB36} & \small 2021 &\small  \centering --  & \small --\\ 
   \hline
      \small  \emph{POC1A} & \small 2021 &\small Bioinformatic analysis revealed it is highly expressed in multiple cancer (including glioma) & \small \cite{Zhao2022} \\ 
   \hline
      \small  \emph{LRRC61} & \small 2021 &\small GBM prognostic marker according to multi-omics bioinformatic analysis.   & \small \cite{Lei2021}\\ 
   \hline
      \small  \emph{PGCP} & \small 2021 &\small  \centering --  & \small --\\ 
   \hline
  \small  \emph{DHRS4} & \small Both &\small Gene knockdown reduces glioma proliferation, invasion, and migration.   & \small \cite{Dai2020a}\\ 
   \hline
  \small  \emph{EEF1A1P9} & \small Both &\small Bioinformatic analysis revealed that this pseudogene predicts survival in glioma   & \small \cite{Wang2019}  \\
   \hline
  \small  \emph{RPL7A} & \small Both &\small Bioinformatic analysis revealed that it is upregulated and related to survival in GBM.  & \small \cite{Wang2020a}\\
   \hline
  \small  \emph{SNAP91} & \small Both &\small  Downregulated and correlated with survival in GBM. & \small \cite{Gao2016a}\\
   \hline
  \small  \emph{EIF3L} & \small Both &\small  \centering --  & \small -- \\
   \hline
  \small  \emph{POM121C} & \small Both &\small  \centering --  & \small -- \\
   \hline
  \small  \emph{GTF2IRD2} & \small Both &\small Role in the neurodevelopmental disorder Williams-Beuren syndrome. & \small \cite{Porter2012}\\
   \hline
  \small  \emph{S100A11} & \small Both &\small Its expression positively correlates with glioma survival.   & \small \cite{Wang2021a} \\
   \hline
  \small  \emph{FAM115C} & \small Both &\small Tumor suppressor associated with prolonged survival in pancreatic cancer. & \small \cite{Saeki2020}\\
   \hline
  \small  \emph{KIF20A} & \small Both &\small Overexpression promotes glioma progression. Downregulation inhibits glioma tumorigenesis.   & \small \cite{Wuang2017,Duan2016}\\
   \hline
  \small  \emph{AURKA} & \small Both &\small   Involved in the self-renewal of in glioma-initiating cells.  & \small \cite{Xia2013} \\
   \hline
  \small  \emph{SGOL1} & \small Both &\small Bioinformatic analysis revealed that downregulation determines better prognosis in glioma.  & \small \cite{Dai2017a} \\
  \hline
  \caption{Literature research has been performed on genes identified by the regularized Cox regression algorithm to have predictive significance. The second column specifies if the dataset leading to the gene selection has been created by considering 2016 or 2021 classifications (or both). If the research resulted in noteworthy information, they are reported with the corresponding reference.
  } \label{tab:literature}
	\end{longtable}
\end{document}